\documentclass[pre,twocolumn,superscriptaddress,letterpaper]{revtex4}
\usepackage{epsfig}

\begin{document}

\title{Generalized Ensemble and Tempering Simulations: A Unified View}

\author{Walter Nadler}
\email{wnadler@mtu.edu}
\affiliation{
Department of Physics,
Michigan Technological University,
1400 Townsend Drive,
Houghton, MI 49931-1295, USA 
}

\author{Ulrich H. E. Hansmann}
\email{hansmann@mtu.edu}
\affiliation{
Department of Physics,
Michigan Technological University,
1400 Townsend Drive,
Houghton, MI 49931-1295, USA 
}
\affiliation{
John-von-Neumann Institute for Computing, 
Forschungszentrum J\"ulich, 
D-52425 J\"ulich, Germany
}

\date{\today}

\begin{abstract}
From the underlying Master equations we derive one-dimensional stochastic processes that describe generalized ensemble simulations as well as tempering (simulated and parallel) simulations. The representations obtained are either in the form of a one-dimensional Fokker-Planck equation or a hopping process on a one-dimensional chain. In particular, we discuss the conditions under which these representations are valid approximate Markovian descriptions of the random walk in order parameter or control parameter space. They allow a unified discussion of the stationary distribution on, as well as of the stationary flow across each space. We demonstrate that optimizing the flow is equivalent to minimizing the first passage time for crossing the space, and discuss the consequences of our results for optimizing simulations. Finally, we point out the limitations of these representations under conditions of broken ergodicity.

\end{abstract}

\maketitle

\section{Introduction}

The effective simulation of complex thermal systems like proteins and glasses is a constant challenge in contemporary computational physics.
Markov chain Monte Carlo simulation techniques for these systems have undergone remarkable advances in the last decades. Two main classes that have evolved are generalized ensemble and parallel tempering methods. 

In the generalized ensemble (GE) approach \cite{HO96g} the goal is to sample the state space of a physical system so that particularly rare but important states, e.g. low energy or barrier states, are encountered frequently. 
A variety of weight functions have been tested, as well as methods for iteratively improving these functions \cite{Berg96}. 

A persistent problem of such simulations is that relaxation is slow due to barriers and bottlenecks, and for a long time it was not clear whether and how they can be controlled by using particular weight functions on the usual order parameter spaces. Parallel tempering (PT) - sometimes also called replica exchange method - promised a way out of this dilemma 
\cite{Geyer1995,HN,H97f}. Here simulations are performed in parallel at different values of a control parameter, most often the temperature. At certain times the current conformations of replicas at neighboring control parameter values are exchanged according to a generalized Metropolis rule. Thereby an individual replica could perform an additional random walk in control parameter space and - due to shorter relaxation times in some control parameter regime - explore state space more evenly.

However, the problem of slow relaxation arises also this time in the form
of a slow and possibly uneven random walk through control parameter space. At least part of this problem is related to finding an efficient discretization of control parameter space.

Increasing the flow through order parameter space in GE sampling as well as through control parameter space in PT was always an incentive. However, usually it was discussed only informally, and only recently Trebst et al. \cite{Trebst2004,Trebst2006a,Trebst2006b, Katzgraber2006} have made an attempt to look at that problem systematically. Instead of concentrating on the stationary distributions that arise from the sampling, they concentrated on the stationary flow across order parameter and control parameter space. In order to optimize the flow they derived weight functions and control parameter discretization schemes.

In this contribution we want to give that approach a more fundamental basis. We will first concentrate on the underlying Master equations describing GE and PT simulations. From them we will derive in a systematic way the one-dimensional stochastic equations that form the basis for a flow analysis in order parameter and control parameter space. These equations will also allow us to investigate connections between flow analysis and another concept describing the dynamics of stochastic processes, the first passage time (FPT). The one-dimensional representations are valid approximations for the underlying simulations only under certain conditions. If they are violated, optimization schemes may still fail. For parallel tempering we will find a criterion from which the validity can be determined.

We will focus in the next section on generalized ensemble sampling, while parallel tempering is the focus of the third section. 
We will close with a discussion of the effects of broken ergodicity on our results and an outlook.

\section{Generalized Ensemble Sampling}

Markov chain Monte Carlo simulations utilize a certain move set in combination 
with an acceptance probability, most often of Metropolis form \cite{Metropolis}, to impose stochastic dynamics on a physical system.
A move from state $s$ to $s'$ is accepted with the probability 
\begin{equation}
p_M(s\to s')=\min\left[1,w(s')/w(s)\right] \quad .
\label{Metropolis}
\end{equation}
The original choice for the weight function $w(s)$ is the thermal or Boltzmann weight 
\begin{equation}
w(s)\propto\exp[-\beta E(s)] \quad ,
\label{BoltzmannWeight}
\end{equation}
resulting in canonical sampling at inverse temperature $\beta=1/k_B T$. However, other choices are possible as well, depending on which particular aspect of state space should be emphasized.
We will assume in the following that the weight function is based only on the energy $E(s)$ of the state $s$, and we will use $w(s)=w\left[E(s)\right]=w(E)$ interchangeably.

In order to get a deeper understanding of simulations based on Eq.~(\ref{Metropolis}) we should look at their description via a Master equation in state space. $P(s,t)$ is the probability to be in state $s$ at time $t$, and its evolution in discrete computer time is given by
\begin{widetext}
\begin{equation}
P(s,t+1) = \sum_{s'\ne s} P(s',t)W_s(s'\to s) + P(s,t) \left[ 1 - \sum_{s'\ne s} W_s(s\to s')\right] \quad ,
\label{MEstate}
\end{equation}
where the sums are over all possible states. The transition probabilities $W(s\to s')$ in this equation are 
\begin{equation}
W_s(s\to s') =
{\psi(s\to s')\over \sum_{s"} \psi(s\to s")} p_M(s\to s') \quad ,
\label{Ws}
\end{equation}
where $\psi(s\to s')$ is the characteristic function of the move set with
$\psi(s\to s')=1$ if the move from $s$ to $s'$ is allowed and zero otherwise.
Provided the move set is balanced and ergodic, the $stationary$ $distribution$ reached by this Markov chain is $P_0(s) \propto w(s)$ because of detailed balance. Note however that in addition to $\psi(s\to s')=\psi(s'\to s)$, for $(s,s')$ with $\psi(s\to s')=1$ the property $\sum_{s"}\psi(s\to s")=\sum_{s"}\psi(s'\to s")$ must be fulfilled.  Particularly the latter property, i.e. every state must connect to the same number of other states, is sometimes overlooked.

Equation~(\ref{MEstate}) is an exact description of the simulation process and it is also the basis for more thorough investigations in the mathematics of Metropolis simulations \cite{Diaconis1996,Diaconis1998,Diaconis2001,Diaconis2004}. However from a physicist's point of view a {\it reduced} description in terms of {\it slow order parameters} is of more interest. Prominent among the order parameters chosen is the energy itself. Using adiabatic elimination of fast degrees of freedom, see the Appendix~\ref{adiabatic}, an approximate  Master equation in energy space can be formulated
\begin{equation}
P(E,t+1) = \sum_{E'\ne E} P(E',t)W_E(E'\to E) + P(E,t) \left[ 1 - \sum_{E'\ne E} W_E(E\to E') \right] \quad .
\label{MEenergy}
\end{equation}
The effective transition probabilities $W_E(E\to E')$ can be derived from Eqs.~(\ref{MEstate} and (\ref{Ws}), and are given in that appendix, too. Note however that, in contrast to Eq.~(\ref{MEstate}), Eq.~(\ref{MEenergy}) is  an approximation, valid only if all other degrees of freedom relaxate much faster than the energy. Nevertheless, even if relaxation orthogonal to the energy is slow, Eq.~(\ref{MEenergy}) can still be viewed as a $Markovian$ approximation to the fully non-Markovian process. Due to the degeneracy of states with energy, the stationary distribution of Eq.~(\ref{MEenergy}) is now 
\begin{equation}
P_0(E) \propto n(E)w(E) \quad . 
\label{nEwE}
\end{equation}
Here, $n(E)$ is the density of states and we have assumed that the 
weight function for the Metropolis algorithm is based on energies only, as mentioned above in the discussion of Eq.~(\ref{Metropolis}).

Coarse graining time and energy leads to a form of the Master equation that is continuous in both variables 
\begin{equation}
{\partial\over \partial t} P(E,t) = \int_{E'} P(E',t)R_E(E'\to E) - P(E,t)\int_{E'} R_E(E\to E') \quad .
\label{MEenergyCont}
\end{equation}
\end{widetext}
where the transition probabilities have been replaced by rates $R_E(E\to E')$. Note that for various systems state space and energy $E$ could have been continuous from the start, i.e. the sums in Eqs.~(\ref{MEstate}) and (\ref{MEenergy}) could have already been integrals.
The continuous form of the Master equation in energy space is now the starting point for our final approximation. If the transition rates $R_E(E\to E')$ are strongly peaked around $E'\approx E$, a second order partial differential equation can be derived from Eq.~(\ref{MEenergyCont}).
by various techniques, e.g. Kramers-Moyal expansion \cite{vanKampen,Risken,Gardiner}. This Fokker-Planck equation \cite{Risken} for $P(E,t)$  is given by
\begin{equation}
{\partial\over \partial t} P(E,t)={\partial\over \partial E} D(E)\left[{\partial\over \partial E}-F(E)\right]P(E,t) \;,
\label{Smoluchowski}
\end{equation}
and we have written it already in a form that
 separates static and dynamic properties. $D(E)$ is the energy-dependent diffusion coefficient that describes the local mobility and $F(E)$ is the drift term which in one dimension can always be derived from a potential, $F(E)=-(d/dE)U(E)$. In particular the stationary distribution of Eq.~(\ref{Smoluchowski}) is fully determined by this potential only,
\begin{equation}
P_0(E)\propto \exp\left[-U(E)\right] \quad .
\end{equation}

It is important to emphasize again that the transition from the Master equation~(\ref{MEenergyCont}) to the Fokker-Planck equation~(\ref{Smoluchowski}) is possible only if the the transition rates $R_E(E\to E')$, i.e. the underlying transition probabilities $W_E(E\to E')$, are {\it sufficiently local} in the energy. Only in that limit the Fokker-Planck equation is an effective description of the more general Eq.~(\ref{MEenergyCont}). Nevertheless, even if there are non-local contributions to the transition rates,  Eq.~(\ref{Smoluchowski}) can be viewed as the best {\it local} approximation to  Eq.~(\ref{MEenergyCont}).

Eq.~(\ref{Smoluchowski}) can be written in a more compact form using the stationary distribution $P_0(E)$, 
\begin{equation}
{\partial\over \partial t} P(E,t)=
\left[{\partial\over \partial E} D(E)P_0(E){\partial\over \partial E} P_0(E)^{-1}\right]P(E,t)\;.
\label{Smoluchowski2}
\end{equation}
In this form  the fact that $P_0(E)$ is the stationary distribution can be seen immediately from the vanishing of the rightmost derivative on the $rhs$ if $P(E,t)$ is replaced by $P_0(E)$.
This equation will be the basis for our further analysis of distribution and flow in energy space. 

The stationary distribution in energy space, $P_0(E)$, is actually the histogram $H(E)$ of the energy distribution that is observed in an actual simulation. It is still given by Eq.~(\ref{nEwE}), i.e. by the Metropolis weight function $w(E)$ multiplied with the density of states $n(E)$. By an appropriate choice of the weight function $any$ histogram can be produced in the simulation. The usual choice of Boltzmann weights leads to the canonical distribution, $H(E) \propto n(E)\exp(-\beta E)$,
while the choice $w(E)\propto 1/n(E) $ leads to a flat histogram $H(E)=const$ \cite{Berg96,BergNeuhaus}. A flat histogram would be most appropriate to describe the properties of the system in question over a wide temperature range with equal accuracy, provided the Monte Carlo error at each energy is the same. Various methods have been discussed to actually obtain approximations to $n(E)$ by iteratively improving simulations \cite{Berg96,WangLandau}. However, all of them are still plagued by the problem that equilibration in the system can be slow, in particular when a wide energy range is considered \cite{Dayal2004,Alder2004}. Moreover, it turned out that -- even if a flat histogram is reached -- the error distribution is not flat at all \cite{condmat,Trebst2004}.

It was the important new step by Trebst et al. \cite{Trebst2004} to look systematically at the flow in energy space in simulations. Instead of monitoring the histogram $H(E)$, corresponding to the stationary distribution, they added a label to the system and changed its value whenever it reached minimal and maximal values in the energy, i.e.  $E_{min}$ and $E_{max}$. By counting
just these labels at each energy $E$, the distributions of systems moving up and down in energy, denoted by $n_{up}(E)$ and $n_{down}(E)$, respectively, can be monitored.
The original histogram is recovered from $H(E)=n_{up}(E) + n_{down}(E)$. However, in this way it is also possible to measure the fraction of systems moving up,
\begin{equation}
f_{up}(E) = {n_{up}(E) \over n_{up}(E) + n_{down}(E) } \quad ,
\label{fEsim}
\end{equation}
and, correspondingly, that of systems moving down in energy, $f_{down}(E)$. Note that $f_{up}(E)+f_{down}(E)=1$. Both distributions actually are the stationary distributions of probability flow in the systems with boundary conditions 
$f_{up}(E_{min})=1,f_{up}(E_{max})=0$ and $f_{down}(E_{min})=0,f_{down}(E_{max})=1$, respectively.

This flow in energy space can now be analyzed using the above Fokker Planck equations. Equations.~(\ref{Smoluchowski}, \ref{Smoluchowski2}) are actually continuity equations for the probability flow,
\begin{equation}
{\partial\over \partial t} P(E,t)= {\partial\over \partial E} J(E,t) \quad ,
\label{continuityE}
\end{equation}
with $J(E,t)$ the probability current.
The current between $E_{min}$ and $E_{max}$ can now be determined as the stationary solution of (\ref{continuityE}),
\begin{equation}
J=
\left[D(E)P_0(E){\partial\over \partial E}P_0(E)^{-1}\right]P_{J}(E)
\equiv const \; ,
\label{flow}
\end{equation}
with $P_{J}(E)$ being the stationary distribution for the flow under the above boundary conditions. Note that the stationary distribution $P_0(E)$ discussed before is actually the solution of Eq.~(\ref{flow}) for zero flow!
Integrating that equation we obtain
\begin{eqnarray}
{P_{J}(E)\over P_0(E)} - {P_{J}(E_{min})\over P_0(E_{min})} = \nonumber \\
J \int_{E_{min}}^E dE' {1\over D(E')P_0(E')} 
\label{stationaryCurrentDistribution}
\end{eqnarray}
Using any of the above boundary conditions the total flow across energy space is then given by
\begin{equation}
|J|=\left<\left[D(E)P_0(E)\right]^{-1}\right>^{-1}
\label{J}
\end{equation}
where we used the notation $<.>=\int_{E_{min}}^{E_{max}}dE.$

In order to optimize the weight function $w(E)$ used for the simulation to reach maximal flow across energy space, Trebst et al. maximized Eq.~(\ref{J})
under the constraint of keeping the distribution $P_0(E)$ normalized, which is done by adding a Lagrange multiplier
\begin{equation}
\frac{\delta}{\delta P_0(E)}
\left[ \left<\left[D(E)P_0(E)\right]^{-1}\right>^{-1} + 
\lambda \left<P_0(E)\right> \right] = 0
\label{dJ}
\end{equation}

Stationary flow is not the only concept that can be used to investigate the stochastic dynamics in a system. Another often used concept is the mean first passage time (FPT) \cite{Redner}.
It is the average time a particle starting at one end of a diffusion space needs to reach the other end for the first time. The total mean first passage time for crossing energy space from $E_{min}$ to $E_{max}$ in both directions,
\begin{equation}
\tau = \tau(E_{min}\to E_{max})+\tau(E_{max}\to E_{min})
\label{tau0}
\end{equation}
can be derived from Eqs~(\ref{Smoluchowski},\ref{Smoluchowski2}) and is given by \cite{SSS81}
\begin{eqnarray}
\tau &=&
\int_{E_{min}}^{E_{max}} dE \frac{1}{D(E)P_0(E)} \int_{E_{min}}^E dE' P_0(E') +
\nonumber \\
& & 
\int_{E_{min}}^{E_{max}} dE \frac{1}{D(E)P_0(E)} \int_E^{E_{max}} dE' P_0(E') 
\nonumber \\
&=&
\left<\left[D(E)P_0(E)\right]^{-1}\right>\left<P_0(E)\right>
\label{tauCont}
\end{eqnarray}
Note that $P_0(E)$ does not have to be normalized here. Minimization of $\tau$ with respect to $P_0(E)$ leads to
\begin{equation}
\frac{\delta}{\delta P_0(E)}
\left[ \left<\left[D(E)P_0(E)\right]^{-1}\right>\left<P_0(E)\right> \right] = 0
\label{dtau}
\end{equation}
Both variational equations, (\ref{dJ}) and (\ref{dtau}), lead to the same solution
\begin{equation}
P_{0,opt} \propto \frac{1}{\sqrt{D(E)}} \quad ,
\label{P0opt}
\end{equation}
which is already derived in \cite{Trebst2004} from Eq.~(\ref{dJ}).
This current-optimized stationary distribution leads to a symmetric form of the Fokker Planck equation,
\begin{equation}
{\partial\over \partial t} P(E,t)=
\left[{\partial\over \partial E} \sqrt{D(E)}
{\partial\over \partial E} \sqrt{D(E)}\right]P(E,t)\;.
\label{Smoluchowski3}
\end{equation}

In order to reach such a current optimized histogram, $H_{opt}(E)$, in an actual simulation, the weight function has to be chosen accordingly.
For this purpose it is necessary to obtain the local diffusion coefficient from a simulation using some initial weight function $w(E)$. 
Differentiating Eq.~(\ref{stationaryCurrentDistribution}), $D(E)$ is obtained from
\begin{equation}
D(E) = \left[P_0(E){d\over dE}{P_{J}(E)\over P_0(E)} \right]^{-1}
= \left[H(E)f'(E)\right]^{-1}
\label{DE}
\end{equation}
where the we used the fact that  
\begin{equation}
f(E) = {P_{J}(E)\over P_0(E)}
\label{fEtheory}
\end{equation}
holds. Both, $f_{up}(E)$ as well as $f_{down}(E)$ can be used for that purpose. However, some smoothening may have to be performed to obtain a smooth derivative, as discussed in \cite{Trebst2004}.
Since $P_{0,opt}(E)=n(E)w_{opt}(E)$, the optimized weight function is given by
\begin{equation}
w_{opt}(E) ={1\over n(E)\sqrt{D(E)} } \quad .
\label{wopt0}
\end{equation}
Using the fact that $n(E)$ is obtained from the actual simulation by $n(E)=H(E)/w(E)$, we arrive at the iteration formula
\begin{equation}
w_{opt}(E) = w(E) \sqrt{\frac{f'(E)}{H(E)}}
\label{wopt}
\end{equation}
At the fixed point of this iteration $f'(E)=H(E)$, leading to $H(E)=1/\sqrt{D(E)}$ as required, see Eq.~(\ref{DE}).

In an actual simulation situation one iteration might not suffice. This is discussed in detail in \cite{Trebst2004,Trebst2006a,Trebst2006b}. 
In the following we will attempt to apply the same approach to tempering simulations.

\section{Parallel Tempering}

Generalized ensemble sampling can be extended by adding movement in {\it control parameter space}. In addition to moves among different states of the system at a particular value of the control parameter, moves along one or more control parameter directions are possible. The motivation  behind such an extension  is that relaxation in some control parameter regime may be faster, thereby facilitating relaxation in the whole $state+control$ $parameter$ $space$ \cite{Geyer1995}.

Usually, the motion along control parameter directions is not made continuous. Instead, a list of monotonically increasing or decreasing values $\beta_n$ is chosen, thereby introducing control parameter hopping. The most commonly used control parameter is temperature, although other choices are possible, too \cite{KH}. 
Simulations at a particular control parameter value are performed using the Metropolis criterion (\ref{Metropolis}) with a weight function $w(\beta,s)$, as before. Although  Boltzmann weights are usually chosen, leading to canonical simulations at each parameter value, in principle any weight function is possible.  

Parallel tempering is the parallellized extension of a method called
simulated tempering which we will sketch only briefly.
In simulated tempering \cite{Lubartsev1992,Marinari1992}, a single instance of the system is simulated only. After certain times, an attempt is made to change the control parameter value. In order to ensure that the stationary distribution at each control parameter value is given by $w(\beta,s)$, the  Metropolis criterion,
\begin{equation}
p_M\left[(\beta,s)\to (\beta',s)\right]=
\min\left(1,{w(\beta',s)\over w(\beta,s)}\right) \quad ,
\label{MetropolisST}
\end{equation}
is used for the acceptance of a control parameter change. In order to ensure appreciable exchange between different
control parameter values using Boltzmann weights (\ref{BoltzmannWeight}), the weight function has to be adjusted by a control parameter dependent function $g(\beta)$,
\begin{equation}
w(\beta,s)= \exp\left[-\beta E(s) + g(\beta)\right] \quad .
\label{STweight}
\end{equation}
This function $g(\beta)$ actually determines the distribution of the single system among the control parameter values and  -- up to an additive constant -- the optimal choice is the free energy $ g(\beta) =\beta F(\beta)$ of the system analyzed \cite{Marinari1992}.
However, the problem of determining $g(\beta)$ has hampered this approach.

This problem is solved in parallel tempering. Here, copies of the system are simulated in parallel at each of the various control parameter values. At certain times {\it exchanges} of replicas with neighboring control parameter values are attempted.
In order to ensure that the stationary distribution at a control parameter value $\beta$ is given by $w(\beta,s)$, the generalized Metropolis criterion 
\begin{equation}
p_M\left[(\beta,s)\to \beta',s')\right]=
\min\left(1,{w(\beta,s')w(\beta',s)\over w(\beta,s)w(\beta',s')}\right)
\label{MetropolisPT}
\end{equation}
has to be used for the acceptance of such an attempt. 
It can be seen easily that any function $g(\beta)$ in the weight function (\ref{STweight}), in particular the one that was necessary to ensure equilibration among control parameter values in simulated tempering,
simply drops out in the parallel form. Thereby, the problem of determining
the free energy in order to optimize the simulation vanishes.
In the case of Boltzmann weights (\ref{MetropolisPT}) reduces to
\begin{equation}
p_M\left[(\beta,s)\to (\beta',s')\right]=
\min\left[1,\exp(\Delta\beta\Delta E\right] \quad
\label{MetropolisPTsimple}
\end{equation}
with $\Delta\beta=\beta'-\beta$ and $\Delta E=E'-E$. 

We will concentrate on temperature hopping in the following and choose
 the list of inverse temperatures $\beta_0>\beta_1>...>\beta_N$. In a parallel implementation
simulations at a particular value $\beta_n$ are usually run on
a particular node of the parallel computer, conveniently labeled $n$.
In order to simplify the notation, we will therefore abbreviate $\beta_n$ by 
$n$ whenever possible and also use "node" synonymously with "control parameter value".

It is sufficient to follow only a single replica through state and control parameter space, since all replicas are equivalent. 
For times between replica exchanges, simulations are performed 
at each node independently, and the time evolution of the distribution function $P\left[(\beta_n,s),t\right]$ at a particular node $n$ is described by the Master equation~(\ref{MEstate}) with the respective transition probabilities determined by the appropriate temperature $\beta_n$. For times $t=m T$, $m=1,2,...$, replica exchange is attempted. For this time step the Master equation in state and temperature space is 
\begin{widetext}
\begin{eqnarray}
P\left[(\beta_n,s),t+1\right]&=&
\sum_{s'} \left\{
P\left[(\beta_{n-1},s'),t\right]W_s\left[(\beta_{n-1},s')\to (\beta_n,s)\right] + P\left[(\beta_{n+1},s'),t\right]W_s\left[(\beta_{n+1},s')\to(\beta_n,s)\right] \right\} + 
\nonumber \\
& &
P\left[(\beta_n,s),t\right]\sum_{s'} \left\{1 -
W_s\left[(\beta_n,s)\to (\beta_{n-1},s')\right] -
W_s\left[(\beta_n,s)\to (\beta_{n+1},s')\right] \right\} \quad .
\label{MEstateT}
\end{eqnarray}
 The transition probabilities are given by
\begin{equation}
W_s\left[(\beta,s)\to (\beta',s')\right] = \frac{1}{\Omega N}
 p_M\left[(\beta,s)\to (\beta',s')\right] \quad
\hbox{for} \quad s\ne s' \quad . 
\label{WPT}
\end{equation}
Here, $\Omega$ is the normalization by state space and reflects the fact that any conformations can be exchanged, which is different from the case of GE where the move set was restricting possible conformation changes, see Eq.~(\ref{Ws}).
$N$ takes into account that just for one random neighboring pair of nodes a replica exchange is attempted at time $t=mT$. However, other strategies are possible, too, that would lead to a different normalization constant.
Due to the exchange of replicas in parallel tempering the transition probabilities are symmetric, i.e. 
\begin{equation}
W_s\left[(\beta,s)\to (\beta',s')\right]=
W_s\left[(\beta',s')\to (\beta,s)\right] \quad .
\label{symmetryWsBeta}
\end{equation}
Elimination of fast degrees of freedom orthogonal to the energy is possible in the same way as it was discussed in the last section. This leaves us with 
\begin{eqnarray}
P\left[(\beta_n,E),t+1\right]&=&
\sum_{E'} \left\{
P\left[(\beta_{n-1},E'),t\right]W_E\left[(\beta_{n-1},E')\to (\beta_n,E)\right] + P\left[(\beta_{n+1},E'),t\right]W_E\left[(\beta_{n+1},E')\to (\beta_n,E)\right] \right\} +
\nonumber \\
& &
P\left[(\beta_n,E),t\right]\sum_{E'} \left\{ 1 -
W_E\left[(\beta_n,E)\to (\beta_{n-1},E')\right] -
W_E\left[(\beta_n,E)\to (\beta_{n+1},E')\right] \right\}
\label{MEenergyBeta}
\end{eqnarray}
The symmetry (\ref{symmetryWsBeta}) naturally leads to a similar symmetry for the transition rates between nodes in reduced, i.e. energy, space,
\begin{equation}
W_E\left[(\beta,E)\to (\beta',E')\right]=W_E\left[(\beta',E')\to (\beta,E)\right] \quad .
\label{symmetryWEBeta}
\end{equation}
In order to finally derive an effective Master equation for hopping in temperature space, we have to additionally assume fast relaxation in energy space, i.e. on times scales $t<T$. This means that at any particular node we assume to have reached the respective equilibrated distribution $P_0(\beta,E)$. Using similar reasoning as in Appendix~\ref{adiabatic} for GE,
this last approximation leads to the final form of the Master equation in temperature space on a coarse grained time scale $t\to t/T\,N$,
\begin{eqnarray}
P(\beta_n,t+1) &=& P(\beta_{n-1},t)W_\beta(\beta_{n-1}\to \beta) + P(\beta_{n+1},t)W_\beta(\beta_{n+1}\to \beta) + 
\nonumber \\
& &
P(\beta_n,t)\left[1 - W_\beta(\beta\to \beta_{n-1}) - W_\beta(\beta\to \beta_{n+1})\right]
\label{MET}
\end{eqnarray}
In a way similar to the derivation of Eq.~(\ref{WE}),
effective transition probabilities can be derived from the equilibrated distributions at a node,
\begin{equation}
W_\beta(\beta\to \beta') = \int dE \int dE' P_0(\beta,E) 
p_M\left[(E,\beta)\to(E',\beta')\right] P_0(\beta',E')  \quad .
\label{Wlimit}
\end{equation}
\end{widetext}
We will discuss in Appendix \ref{AProb} the properties of these effective transition
probabilities in particular situations. Finally, the symmetry of the transition probabilities
\begin{equation}
W_\beta\left(\beta\to \beta'\right)=W_\beta\left(\beta'\to \beta\right) 
\end{equation}
holds analogously here, too.

Due to this last property, the stationary distribution for PT can be derived easily from Eq.~(\ref{MET})
and is given simply by
\begin{equation}
P_0(\beta) = const  \quad ,
\label{P0discrete}
\end{equation}
i.e. in an equilibrated simulation every single replica appears on each control parameter node with the same probability, $1/(N+1)$ with $N+1$ being the number of nodes. This result is an important simplification of the situation over the case of simulated annealing and over GE, and is due to the construction of replica exchange.

Flow between the control parameter nodes can now be analyzed, too. Here, replicas reaching node $0$ or $N$ are labelled and the respective distributions over the nodes can be monitored.
Using the same notation as before, $n_{up}(i)$ being the number of replicas at node $i$ that came from node $0$, and $n_{down}(i)$ being the number of replicas at node $i$ that came from node $N$, we can measure the fraction of replicas moving up
\begin{equation}
f_{up}(i) = \frac{n_{up}(i)}{n_{up}(i)+n_{down}(i)} \quad ,
\end{equation}
and a corresponding quantity for those moving down, $f_{down}(i)$.
Both distributions are stationary distributions of probability
flow between temperature nodes, with boundary conditions $f_{up}(0)=1,f_{up}(N)=0$ and $f_{down}(0)=0,f_{down}(N)=1$, respectively.
As before, they can be analyzed by looking at the
stationary solution of the underlying stochastic equation (\ref{MET}).
It can be written as
\begin{equation}
P(\beta_n,t+1) - P(\beta_n,t) = J(\beta_n,t) - J(\beta_{n-1},t)
\label{continuityEdiscrete}
\end{equation}
which is the discrete form of the continuity equation (\ref{continuityE}).
The discrete case current $J(\beta_n,t)$ is given by
\begin{eqnarray}
J(\beta_n,t) &=& P(\beta_{n+1},t)W_\beta(\beta_{n+1}\to \beta_n) - \nonumber\\
& & P(\beta_n,t)W_\beta(\beta_n\to \beta_{n+1}) \quad .
\label{Jdiscrete}
\end{eqnarray}
Consequently,  the stationary current $J$ is determined from 
\begin{eqnarray}
J &=& P_J(\beta_{n+1})W_\beta(\beta_{n+1}\to \beta_n) - \nonumber\\
& & P_J(\beta_n)W_\beta(\beta_n\to \beta_{n+1}) \nonumber\\
&=& const \quad ,
\label{JdiscreteStationary}
\end{eqnarray}
with $P_J(\beta_n)$ being the stationary distribution for flow under the above
boundary conditions.
Taking into account the symmetry properties of the transition probabilities, the stationary distribution for constant current between nodes $\beta_0$ and $\beta_N$
is given by
\begin{equation}
P_{J}(\beta_n) = J^{-1} \sum_{i=0}^{n-1} {1\over W_\beta\left(\beta_i\to \beta_{i+1}\right)}
\label{PJdiscrete}
\end{equation}
with the current
\begin{equation}
J = \left[\sum_{i=0}^{N-1}{1\over W_\beta\left(\beta_i\to \beta_{i+1}\right)}\right]^{-1}
\label{J_PT}
\end{equation}
Due to the simple form of the stationary distribution among nodes (\ref{P0discrete}), the analytic forms for $P_J(i)$ and $J$ are considerably simpler than for GE.

Again, as before with GE, a concept different from stationary flow can also be analyzed, the total mean first passage time to cross the network of nodes. For the general hopping process (\ref{MET}), this is given by \cite{Gardiner,vanKampen}
\begin{eqnarray}
\tau &=& \tau(0\to N) +\tau(N\to 0) \nonumber \\
&=&
\sum_{i=0}^{N-1}{1\over P_0(\beta_i) W_\beta\left(\beta_i\to \beta_{i+1}\right)}
\sum_{j=0}^i P_0(\beta_i) +
\nonumber \\
& &
\sum_{i=1}^{N}{1\over P_0(\beta_i) W_\beta\left(\beta_{i}\to\beta_{i-1}\right)}
\sum_{j=i}^{N} P_0(\beta_i) 
\nonumber \\
&=& \left(\sum_{i=0}^{N-1}{1\over P_0(\beta_i) 
W_\beta\left(\beta_i\to \beta_{i+1}\right)} \right)
\left(\sum_{i=0}^N P_0(\beta_i) \right) \quad .
\label{tauDiscrete}
\end{eqnarray}
This final result is practically a discrete version of Eq.~(\ref{tauCont}). Taking into account that the stationary distribution 
in PT is constant, we finally obtain a result that is just the inverse of Eq.~(\ref{J_PT}),
\begin{equation}
\tau = \sum_{i=0}^{N-1}{1\over 
W_\beta\left(\beta_i\to \beta_{i+1}\right)} \quad.
\label{tauDiscrete1}
\end{equation}

Here is an important difference to GE. There the local diffusion coefficient was fixed by the move set and - as it turned out in actual simulations - mostly independent from the chosen weight function. The stationary distribution, however, was free to be chosen by varying the weight function.

In the case of PT this is exactly the opposite. Here the stationary distribution is fixed by construction of the replica exchange process. However, by adjusting the control parameter intervals, the transition probabilities can be chosen relatively freely. Hence, we are interested in obtaining the optimal distribution of transition probabilities that maximizes the flow across the control parameter space. Since we are mainly interested in local variations of the optimized transition probabilities, i.e. in deviations from an average value (that will be actually determined afterwards), we keep the average transition probability constant via a Lagrangian multiplier, i.e.
\begin{eqnarray}
{\delta\over\delta W_\beta(j\to j+1) } 
\left[\left(\sum_{i=0}^{N-1}{1\over W_\beta\left(\beta_i\to \beta_{i+1}\right)}\right)^{-1} \right. + & \nonumber \\
\left. \lambda \sum_{i=0}^{N-1} W_\beta\left(\beta_i\to \beta_{i+1}\right) \right) =  0 & \quad .
\end{eqnarray}
An equivalent equation results for optimizing $\tau$.

It is easily seen that optimizing the current as well as the mean first passage time simply gives a constant transition probability between neighboring nodes the whole range of control parameter values,
\begin{equation}
W^{opt} = const
\label{Wconst}
\end{equation}
as optimal solution. Consequently, from (\ref{PJdiscrete}) we can conclude that the optimal flow distribution among the nodes
 is linear in the node number.
\begin{equation}
P_{J}(\beta_n) = n/N \quad .
\label{PnN}
\end{equation}
Therefore, the temperature spacing is optimal if such a flow distribution $together$ with constant transition probabilities can be obtained in an actual simulation. The linear dependence of the flow distribution on the node number, Eq.~(\ref{PnN}), was obtained for PT already in \cite{Trebst2006a, Trebst2006b,Katzgraber2006} by mapping PT onto the Fokker-Planck equation for GE, Eq~(\ref{Smoluchowski}), and assuming a particular temperature dependence for the local diffusion coefficient. Here however, we see that it follows directly from the hopping-description of PT. In addition, we obtain its equivalence to constant transition probabilities, Eq.~(\ref{Wconst}).

The iteration scheme used in Refs.~\onlinecite{Trebst2006a, Trebst2006b, Katzgraber2006} for assigning temperatures to nodes appeared to exhibit fast convergence to the optimal behavior of \ref{PnN}. We can rephrase it here without having to recur to some intermediate local diffusivity (and we stick to our use of inverse temperatures as example control parameters): \break
(i) a particular set of control parameters $\beta_0>\beta_1>...>{beta_N-1}>\beta_N$ gives rise to a flow distribution 
$f_{up}(0)=1\ge f_{up}(1)\ge ...\ge f_{up}(N-1)\ge f_{up}(N)=0$; \break
(ii) these latter values give rise to stepwise defined function $g[f]$, with $g[f(i)]=\beta_i$, in particular $g[1]=\beta_0$ and $g[1]=\beta_N$, and linear interpolation in between these values; \break
(iii) the new control parameter values are determined from this function by $\beta_i'=g[i/N], i=1,...,N-1$, keeping $\beta_0$ and $\beta_N$ fixed.\break
This procedure is actually illustrated quite nicely in Fig.~2 of Ref.~\onlinecite{Trebst2006a}, and we can refrain from repeating it here.

It is important to note at this point that the above results, in particular the $equivalence$ of constant transition probabilities and a flow distribution that is linear in the node number, depend on the validity of the underlying one-dimensional representation
of the simulation process. It turns out that, while the flow distribution Eq.~\ref{PnN} is readily attainable in actual simulations, this is usually not accompanied  by acceptance probabilities that are constant over the whole system \cite{Trebst2006a, Trebst2006b,Katzgraber2006}.
In the following final section we will discuss possible reasons for such a discrepancy.

\begin{figure}
\epsfig{file=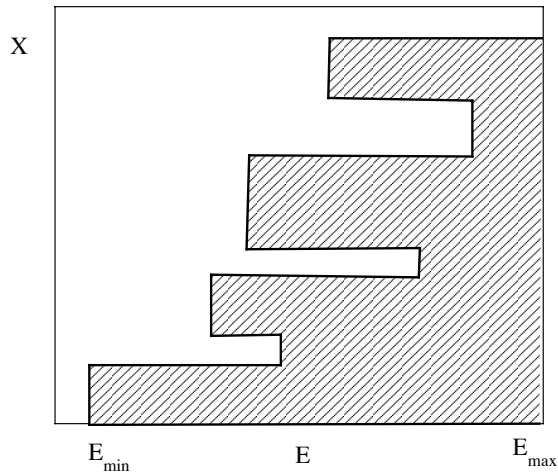,width=8cm}
\caption{ Sketch of state space for generalized ensemble sampling (GE) in the case of broken ergodicity; X denotes any degree of freedom orthogonal to the control parameter (energy).
\label{GE_comb}
}
\end{figure}

\begin{figure}
\epsfig{file=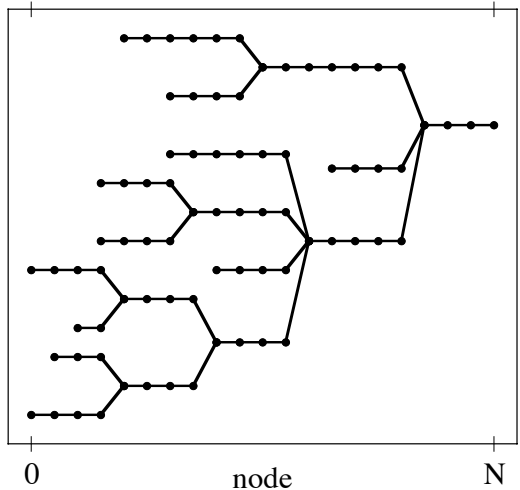,width=8cm}
\caption{ Sketch of bi- and multi-furcations in the case broken ergodicity for parallel tempering (PT); for certain nodes the system partitions into several disjoint free energy wells.
\label{PT_bifurcation}}
\end{figure}

\section{Summary, Discussion, and Outlook}

The goal of most recent advances in Markov chain Monte Carlo sampling is to analyze and increase the flow through state space. To do so, heuristic equations have been used to describe the flow in reduced state space along a slow order parameter, e.g. the energy, in generalized ensemble sampling (GE) and among nodes performing simulations at various control parameter values in parallel tempering (PT) \cite{Trebst2004,Trebst2006a,Trebst2006b}. In this contribution we have derived such one-dimensional stochastic equations for GE and PT sampling from the underlying Master equations. Using
these stochastic equations, weight functions for GE and strategies for
finding optimal control parameter values for PT can be devised that optimize
the flow through order parameter and control parameter space, respectively.
We have also demonstrated that optimization of flow is equivalent to minimizing the first passage time to cross the system.

All considerations in the previous sections were based on the assumptions that 
the Fokker-Planck (\ref{Smoluchowski}) or hopping (\ref{MET}) equations are a valid Markovian representation of the underlying more complex dynamics. That, however, is true only if the approximations discussed there apply. 

In the case of GE these approximations are that relaxations in the degrees of freedom orthogonal to the energy are fast, together with the locality of transitions. 
Since the ultimate goal of deriving Eq~(\ref{Smoluchowski}) is the optimization of flow through state space, a violation of the latter condition is not detrimental. Non-locality in the energy of the move set usually leads to faster relaxation since state space is connected more densely.
This is for example one reason for the success of the Swendsen-Wang algorithm
\cite{SwendsenWang}. Although in such a situation Eq.~(\ref{Smoluchowski}) may not be able to capture the full dynamics correctly, it nevertheless is still able to identify local bottlenecks. Optimizing the flow according to the methods discussed can handle these local bottlenecks in the transition between neighboring energy values.

However, slow relaxation orthogonal to the energy leads to a more complicated situation. Now it may not be possible anymore to reach all values of the additional degrees of freedom by moves local in the energy. Instead, detours via other - usually high energy - areas of the state space have to be performed. This leads to the comb-like structure of the accessible state space sketched in Fig.~\ref{GE_comb}. It describes the situation that free energy basins at constant energy are disconnected. 

Actually, it is this feature of state space partitioning that lead us to the requirement of large flow between low and high energy areas of the state space in the first place. Only for a large flow the "teeth" of the comb-like structure in Fig.~\ref{GE_comb} can be sampled adequately. Nevertheless, it also leads to the situation that the effective one-dimensional Fokker-Planck equation (\ref{Smoluchowski}) is only the best Markovian one-dimensional approximation of an underlying effectively higher-dimensional process.

This situation is even clearer in the case of PT. If the relaxation at a particular control parameter value is faster than the time scale of hopping in control parameter space, then the requirements for the analysis performed in the previous section are fulfilled. However, if that is not the case the state space at such a node partitions into disjoint free energy basins that do not communicate. Viewed over the whole control parameter range, we are dealing with a hierarchical network of free energy basins as sketched in Fig~\ref{PT_bifurcation}. Such a situation has been aptly termed {\it broken ergodicity} \cite{Palmer1982,Stein1995} and was discussed in the field of glassy dynamics several years ago.

In principle, the topology of the tree-like control+state space depends on the time scale of the control parameter hopping. If relaxation is possible at all nodes, then no bifurcations occur and the system is just a one-dimensional hopping chain as it was analyzed in the previous section. However, this is true only in the limit of infinite time between replica exchange steps, $T\to\infty$. Practically the topology of the branching will be the same over a wide range of time scales and only the position of the branching nodes may vary.

For a truly one-dimensional system the optimized transition probabilities are constant and -- equivalently -- the optimal flow stationary distribution is $f(n)\propto n$, Eq.~(\ref{PnN}). Under conditions of broken ergodicity, the situation is more complicated.
Now, several transition rates between neighboring nodes may have to be taken into account, describing exchange along the different branches depicted in Fig.~\ref{PT_bifurcation}. Moreover, {\it observed} acceptance rates are weighted averages of these various transition probabilities. Therefore, the equivalence of constant observed acceptance ratios to a flow distribution linear in the node number may no longer hold. This discrepancy has been observed already in Refs.~\onlinecite{Trebst2006a, Trebst2006b,Katzgraber2006}. 
While satisfying Eq.~(\ref{PnN}) was possible by an appropriate choice of node temperatures, constant acceptance rates were not obtained concomitantly. Such a result can be used as a clear signature of broken ergodicity occurring in PT. In contrast to GE,
where such a clear criterion is not available yet, PT has a particular advantage here.

Naturally, the question arises how to optimize flow in PT under conditions of broken ergodicity. Control via the choice of node temperatures is somewhat limited in such a situation, since changes may affect different transition probabilities differently. 
Nevertheless, even in such a  situation the choice of a flow distribution $f(n)\propto n$ still assures that the flow of replicas along the main branch, i.e. between the lowest and the highest temperature node, is still optimal. In contrast, the effect of making apparent acceptance ratios constant -- if possible at all -- is unclear and depends on a knowledge of the particular structure of ergodicity breaking. In the final analysis, however, optimizing flow under such conditions means that, in addition to the flow between the lowest and the highest node, also flow among side branches has to be be considered. In order to assure that flow among all side branches is optimized, too, a more detailed flow analysis, i.e. determining the flow matrix between all individual nodes, would have to be performed.

We believe that an additional advantage of PT is that such branching situations can be analyzed directly, without resorting to an actual system. In a way, it will be possible to {\it simulate the simulations} to investigate possible flow behaviors. By analyzing the Master equations modelling the hierarchical broken ergodicity networks, conclusions about the behavior of actual simulations can be drawn \cite{NadlerXXXX} Thereby, the present results open the way to investigate the effects of broken ergodicity in GE and PT in a more systematic way.

\begin{acknowledgments}
It is a pleasure to thank P. Grassberger and S. Trebst for a critical reading of the manuscript and the referees for posing stimulating questions that helped to improve the presentation. This research was supported by NSF-grant No. CHE-0313618.
\end{acknowledgments}

\vskip 1truecm

\appendix

\begin{widetext}

\section{Adiabatic elimination of fast degrees of freedom}
\label{adiabatic}

As a first step  to eliminate the fast degrees of freedom in Eq.~(\ref{MEstate}), we have to single out the slow degree of freedom by appropriate labelling. Therefore, we replace the state label $s$ by a more detailed one that consists of the energy $E$, i.e. the slow degree of freedom, and an additional label $s_E$ that designates all microstates with energy $E$, $s\equiv(E,s_E)$. Equation~(\ref{MEstate}) is then replaced by
\begin{eqnarray}
P\left[(E,s_E),t+1\right] &=& 
\sum_{E'\ne E}\sum_{s_{E'}} P\left[(E',s_{E'}),t\right] W_s\left[(E',s_{E'})\to(E,s_{E})\right] +  \nonumber \\
& &
P\left[(E,s_E),t\right] \left\{ 1 - \sum_{E'\ne E}\sum_{s_{E'}} W_s\left[(E,s_{E})\to(E',s_{E'})\right] \right\} \quad , 
\label{MstateDetailled}
\end{eqnarray}
where the sums are over all possible energies $E$ and states $s_E$ for each energy. If relaxation among the microstates for a particular energy is fast, each of these states will assume the same probability, and we can approximate 
\begin{equation}
P\left[(E,s_E),t\right] \approx \frac{1}{N(E)}P(E,t) \quad ,
\label{Papprox}
\end{equation}
with $N(E)$ being the number of states, i.e. $\sum_{s_E} = N(E)$, introduced to ensure correct normalization, $\sum_E P(E,t)=1$.
This approximation is the crucial step, since it allows us to sum over all microstates $s_E$ for each energy in Eq.~(\ref{MstateDetailled}). After carrying out such a summation, we arrive at the Master equation for the energy, Eq.~(\ref{MEenergy}), with the transition probabilities $W_E(E\to E')$ given by
\begin{equation}
W_E(E\to E') = \frac{1}{N(E)} \sum_{s_{E}} \sum_{s_{E'}} W_s\left[(E,s_{E})\to(E',s_{E'})\right]  \quad .
\label{WE}
\end{equation}
Note the asymmetry with respect to $E$ and $E'$ that is due to the non-constant density of states.

A more rigorous and systematic treatment, which would also allow the derivation of corrections, would involve projection operator techniques as they are used, e. g.. in Chap.~6.4 of Ref.~\onlinecite{Gardiner}. However, since the above approach suffices for our purposes, we refrain from embarking on such a more detailed elaboration.

\end{widetext}

\section{Transition probabilities for parallel tempering}
\label{AProb}

In GE, a general analysis of the effective transition probabilities $W(E\to E')$ is not easy since they depend strongly on the -- usually unknown  -- distributions orthogonal to the energy, combined with a possibly sparse move set. In contrast to GE, PT allows more insight into the effective transition probabilities that govern the temperature hopping. Since transitions are possible between any energy values, the complications due to a particular sparse move set do not arise. The only requirement is the assumption of equilibration at each temperature. In this limit, we can calculate the effective transition probability by
\begin{widetext}
\begin{equation}
W(\beta\to \beta') = \int dE \int dE' P_0(\beta,E) 
p_M\left[(E,\beta)\to(E',\beta')\right] P_0(\beta',E') \quad ,
\label{WlimitAppendix}
\end{equation}
with $P_0(\beta,E)$ being the equilibrated distribution at $\beta$ and $p_M$ given by Eq.~(\ref{MetropolisPT}).
\end{widetext}

There have been several approaches to evaluate that formula. Predescu et al. \cite{Predescu2004} and Kofke \cite{Kofke2004} emphasize the importance of taking into account the asymmetry of an actual distribution, having a low energy cutoff and an exponential tail at high energies. Nevertheless, Kone and Kofke \cite{Kone2005} later use an approximation based on a Gaussian approximation, i.e. symmetric without cutoff and non-exponential tail, together with the assumption of constant specific heat over the entire range. All authors limit themselves to unimodal distributions in their analysis and do not question the peak is quadratic in energy.  

However, the distributions change dramatically at critical values of the control parameter, i.e. at first and second order phase transitions. While at second order phase transitions the functional form of the peak changes from quadratic to quartic, at first order phase transitions the energy distributions become bimodal. With respect to the distribution tails, on the other hand, since the goal is anyhow to optimize the transition probabilities, one should try to avoid  control parameter intervals so large that the explicit structure of the tails become relevant.
Moreover, Eq.~(\ref{WlimitAppendix}) is anyhow only valid in the limit of fast relaxation at a node. So it is from the outset only an approximation to the actually observed transition rate.  Since, in order to optimize the flow, large values of the transition probabilities are sought for,
a quantitative analysis makes sense only for the cases where the overlap is appreciable, i.e. for small temperature differences.

We therefore add here our approach to evaluate (\ref{WlimitAppendix}) using
the first order approximations to these distributions, i.e. Gaussians,
\begin{equation}
P_0(\beta,E) \propto 
\exp\left[-{\left[E-{\overline E(\beta)}\right]^2\over 2 \sigma^2(\beta)}\right]
\end{equation}
with ${\overline E(\beta)}$ the average energy and 
$\sigma^2(\beta)={\overline {\left[E-{\overline E(\beta)}\right]^2}}$ the energy fluctuations at $\beta$. However, we avoid the unrealistic and very limiting assumption of a constant specific heat \cite{Kone2005}.

Assuming $\beta<\beta'$, using a step function approximation of error functions that result from the inner integrals, and performing a symmetric evaluation we obtain 
\begin{widetext}
\begin{eqnarray}
W(\beta\to \beta') &\approx& \frac{1}{4}
\exp\left[\Delta\beta\Delta E + 
\frac{1}{2}\Delta\beta^2\left(\sigma+\sigma'\right)\right]
\left[2+
{\rm erf}\left({\Delta E
+\Delta\beta(\sigma^2+\sigma'^2)\over\sqrt{2}\sigma}\right) +
{\rm erf}\left({\Delta E
+\Delta\beta(\sigma^2+\sigma'^2)\over\sqrt{2}\sigma'}\right)
\right] +
\nonumber \\
& &
 \frac{1}{4} \left[
{\rm erfc}\left({\Delta E\over\sqrt{2}\sigma}\right) +
{\rm erfc}\left({\Delta E\over\sqrt{2}\sigma'}\right) \right] \quad ,
\label{Wapprox}
\end{eqnarray}
\end{widetext}
where we have used the abbreviations  $\Delta \beta=\beta-\beta'$ and
$\Delta E = {\overline E(\beta)}-{\overline E(\beta')}$.
We can now employ the fact that the specific heat is given by
\begin{equation}
c = {d\over dT} {\overline E(T)} = -\beta^2 {d\over d\beta} {\overline E(\beta)} \quad ,
\end{equation}
as well as by
\begin{equation}
c = \beta^2 \overline {\left[E-\overline E(\beta)\right]^2}= 
\beta^2 \sigma^2(\beta)
\end{equation}
to obtain a relationship between the derivative of the average energy and the energy fluctuations
\begin{equation}
 - {d\over d\beta} {\overline E(\beta)} =
 \overline {\left[E-\overline E(\beta)\right]^2} \quad .
\end{equation}
Using
\begin{equation}
{\overline E}'(\beta) \equiv 
{d\over d\beta} {\overline E(\beta)} = 
-\sigma^2(\beta) \quad ,
\end{equation}
this relation can be employed to approximate the difference of the average energies by
\begin{eqnarray}
\Delta E &=& {\overline E}(\beta)-{\overline E}(\beta') \nonumber \\
&\approx& \frac{1}{2}
\left({\overline E}'(\beta)+{\overline E}'(\beta')\right) \Delta\beta
\nonumber \\
&=& -\frac{1}{2}\left(\sigma^2+\sigma'^2\right)\Delta\beta 
\end{eqnarray}
Note that we have assumed $\beta<\beta'$, i.e. $\Delta \beta <0$ and, consequently, $\Delta E >0$ in the evaluation of Eq.~(\ref{WlimitAppendix}). This result shows that the exponent in Eq.~(\ref{Wapprox}) cancels and, using ${\rm erf}(-x)=-{\rm erf}(x)$, we have the final approximate expression for the transition probability
\begin{eqnarray}
W(\beta\to\beta') &\approx&
 \frac{1}{2} \left[
{\rm erfc}\left({\Delta E\over\sqrt{2}\sigma}\right) +
{\rm erfc}\left({\Delta E\over\sqrt{2}\sigma'}\right) \right]
\nonumber \\
&=&
\left[
{\rm erfc}\left({|\Delta\beta|(\sigma^2+\sigma'^2)\over 2\sqrt{2}\sigma}\right) + \right. \nonumber \\
& & \left.
{\rm erfc}\left({|\Delta\beta|(\sigma^2+\sigma'^2)\over 2\sqrt{2}\sigma'}\right) \right]  \quad .
\label{WapproxFinal}
\end{eqnarray}

For small values of ${\Delta E/\sigma}$ this can be further approximated to
\begin{eqnarray}
W(\beta\to\beta') &\approx& 1-\frac{1}{\sqrt{2\pi}}
\left({\Delta E\over\sqrt{2}\sigma}+{\Delta E\over\sqrt{2}\sigma'}\right) +
\nonumber \\
 & & \frac{1}{6\sqrt{2\pi}} \left[
\left({\Delta E\over\sqrt{2}\sigma}\right)^3+
\left({\Delta E\over\sqrt{2}\sigma'}\right)^3 \right] +\nonumber \\
& & O(\Delta E^5) 
\quad .
\label{WapproxLin}
\end{eqnarray}

\end{document}